\documentstyle[11pt,newpasp,twoside,psfig]{article}
\begin{document}

\title{Disk-Driven Outflows in AGNs}
\author{Arieh K\"onigl}
\affil{Department of Astronomy \& Astrophysics, University of
Chicago, 5640 S. Ellis Ave., Chicago, IL 60637, U.S.A.}

\setcounter{page}{111}
\index{Koenigl, A.}

\begin{abstract}
 Analysis of spectral absorption features has led to the
identification of several distinct outflow components in
AGNs. The outflowing gas is evidently photoionized by
the nuclear continuum source and originates in
the accretion flow toward the central black hole. The most
likely driving mechanisms are continuum and line radiation
pressure and magnetic stresses. The theoretical modeling of
these outflows involves such issues as: (1) Which of the above
mechanisms actually contributes in each case? (2) How is the gas
uplifted from the underlying accretion disk? (3) How can the
intense central continuum radiation be shielded to allow
efficient radiative driving? (4) Is the outflow continuous or clumpy, and,
if clumpy, what is the nature and dynamical state of
the ``clouds''? This review summarizes recent theoretical
and observational results that bear on these questions and
outlines prospects for further progress.
\end{abstract}

\section{Observational Background}
Recent observational work has strengthened the evidence for
the common existence of gas outflows from the centers of active
galaxies. In the case of Seyfert 1 galaxies, high-resolution UV
and X-ray spectroscopy (involving, in particular, HST, FUSE,
Chandra, and XMM/Newton) has pointed to the presence of gas
outflowing\footnote{The outflow speeds are inferred
from measurements of line blueshifts relative to the systemic
velocity of the galaxy, and are typically significantly
smaller than the line widths of the BELR gas ($\la 10^4\, {\rm
km\, s^{-1}}$). The latter have been attributed to either rotation, turbulence, or
the effect of electron scattering.}  at $\sim 10^2-10^3\, {\rm km\, s^{-1}}$ and most
likely originating on the scale of the broad emission-line
region (BELR) --- $R_{\rm BELR}\approx 0.1\, L_{45}^{0.7}\, {\rm
pc}$ (where $L_{45}$ denotes the optical luminosity in units
of $10^{45}\, {\rm ergs\, s^{-1}}$; Kaspi et al. 2000).

The outflowing gas in Sy 1 galaxies is often seen in
absorption and is inferred to have a global covering factor
$\ga 0.5$ (e.g., Crenshaw et al. 1999). At high resolution
the absorbing gas separates into distinct kinematic components
that are characterized by a range of physical properties (such as the value of the
ionization parameter $U\equiv n_\nu/n_{\rm H} \propto L_{\rm
ion}/n_{\rm H}r^2$, the ratio of the number densities of
ionizing photons and hydrogen nuclei) and of FWHM widths (up to
$\sim 400\, {\rm km\, s^{-1}}$ in the Crenshaw et al. 1999 sample, indicative of macroscopic
motions). Although the strongest absorption components
evidently lie completely outside of the BELR, the high
line-of-sight covering factors derived in certain cases
suggest that at least some of the absorbing gas lies
close to the nucleus. This conclusion is supported by direct
estimates of the density in a number of sources, which indicate that
the respective absorbing component must lie at a distance of
less than a fraction of a parsec from the continuum source
(e.g., Netzer et al. 2002; Gabel et al. 2002). The effect of the partially
ionized ``warm absorber'' gas is manifested in both the UV and
X-ray spectra, and there is evidence that at least some of the UV and X-ray
components overlap. An important issue, not yet fully resolved
in most cases, is the relationship between the different kinematic
and ionization components that are identified in a given source:
are they best described as a single-phase or a multiphase outflow?

Spectral variability studies can help answer the question of
whether the flow is uniform or clumpy. Typical variability time
scales are months to years, and, although in some cases the origin of the variability is
apparently a change in the irradiating continuum (e.g., Kraemer
et al. 2002a), in other cases the preferred explanation is a
change in the column density of the absorbing gas (e.g.,
Crenshaw \& Kraemer 1999): this can be attributed to the transit or evolution of density
inhomogeneities that intercept our line of sight.

Measurements of variability in P Cygni profiles of UV lines have been
interpreted as evidence for a systematic acceleration of the
BELR gas in the Sy 1 galaxy NGC 3516 (Hutchings et
al. 2001). Inferences from X-ray spectroscopy of an increase in
the outflow speed with decreasing degree of ionization in a number of sources
(e.g., Sako et al. 2001; Kaastra et al. 2002) are consistent with
this picture. Evidence for a systematic acceleration has also
been inferred for the narrow emission-line region (NELR) gas in
both Sy 1 (Crenshaw et al. 2000) and Sy 2 (Crenshaw \&
Kraemer 2000) galaxies over distances $\ga 10^2\, {\rm pc}$.

Broad absorption-line QSOs (BALQSOs) are another class of AGNs
where there is strong evidence for distinct outflow
components. Typical speeds are $\sim 10^3-10^4\, {\rm km\,
s^{-1}}$,  but values of up to $\sim 0.1\, c$ have been
inferred. Similarly to the warm-absorber component in Sy 1
galaxies, at least some of this gas evidently lies outside of
the BELR, and the covering factor at the source is $\ga 0.3$
(e.g., Goodrich 1997). However, in contrast with Sy 1 galaxies,
which are typically viewed at a comparatively small angle to the
symmetry axis, BALR outflows may be observed at a relatively
large angle to the axis (see \S 2). The so-called ``associated''
QSO absorption systems (e.g., Richards et al. 1999), which are blueshifted relative to the
source by up to $\sim 0.1\, c$, may potentially originate in the
QSO itself just like BALQSO outflows. This possibility has
proven hard to verify because of the difficulty in pinning down
the distance from the QSO, which might be $\gg R_{\rm BELR}$.

Another potentially important ingredient is dust. The presence
of dust within the outflow has been inferred in some warm absorbers
and BALQSOs. There is also a
direct indication from spectropolarimetry in the Sy galaxy
NGC 1622 for dust moving toward us at a speed of $\sim 10^3\, {\rm km\,
s^{-1}}$ (Goodrich 1989).

\section{Association of Outflows with Disks}

The BELR emission is characterized by single-peaked lines and a
variability pattern in which a change in the continuum flux
produces an earlier response in the red wing of a line like
H$\beta$ than in the blue wing. These characteristics are most
naturally interpreted in terms of a rapidly accelerated outflow
from a rotationally supported disk (e.g., Chiang \& Murray
1996). This interpretation applies to the two main proposed driving
mechanisms of disk outflows: line driving by the disk
radiation (e.g., Murray \& Chiang 1997) and centrifugal driving
along open magnetic field lines that thread the disk (e.g.,
Bottorff et al. 1997). 

A disk-like geometry for the BELR has been inferred from an
observed correlation between the peak line width and the
radio-axis inclination to the line of sight in radio-loud QSOs
(e.g., Vestergaard et al. 2000; Hough et al. 2002). An
equatorial disk geometry has also been indicated for BALR
outflows by spectropolarimetric observations of BALQSOs (e.g.,
Goodrich \& Miller 1995; Cohen et al. 1995).

\section{Theoretical Framework}

A broadly consistent picture of the BELR and BALR in terms of
disk outflows can be summarized as follows:

\noindent
${\bullet}$ The BELR gas (or at least its high-ionization componet) in
Seyfert galaxies and QSOs corresponds to a disk-driven outflow
that is photoionized by the central continuum radiation.

\noindent
${\bullet}$ The BALR outflow is BALQSOs likely has a dominant
contribution from the radiation pressure force exerted by the
central continuum, which accelerates the gas along comparatively
low-latitude trajectories.

\noindent
${\bullet}$ The photoionized gas at the base of the outflow can naturally
account for the X-ray and UV warm-absorber component. The X-rays
are typically absorbed by a larger column than the UV radiation
This can be attributed to
clumpiness (the UV absorption occurs in low-filling-factor
``clouds'') or to the UV absorption occurring predominantly in
the outer, less strongly ionized regions of the photoionized
nuclear gas. 

\noindent
${\bullet}$ The partially ionized gas likely also contributes to the
X-ray absorption in BALQSOs. The comparatively rapid upward acceleration of
the disk outflow results in a vertical density stratification that
gives rise to a systematic decrease in the absorbing column with
increasing latitude. This explains why BALQSOs, which are
evidently viewed close to the disk plane, have low X-ray fluxes
that in most cases can be attributed to a large absorbing
column (e.g., Brandt et al. 2000; Green et al. 2001). Part of the
absorption in BALQSOs and Sy 2 galaxies may occur in a
dusty, molecular ``torus'' located outside the BELR. This proverbial
``torus,'' which is the basis of the unification scheme of type
1 and type 2 sources (either Seyferts or QSOs; e.g., Antonucci 1993)
can be associated with the dusty outer region (beyond the
dust sublimation radius $R_{\rm subl}$) of the disk outflow
(K\"onigl \& Kartje 1994).\footnote{The sublimation radius can
be naturally identified with the outer boundary of the BELR
(Netzer \& Laor 1993).} Some dust absorption in Sy 2
galaxies evidently originates on larger scales ($\ga 10-10^2\,
{\rm pc}$), probably in molecular clouds 
located in the plane of the host galaxy (e.g., Crenshaw \&
Kraemer 2001; Risaliti et al. 2002), but the bulk of the
absorption in Compton-thick sources likely arises in the more
compact ``torus'' (e.g., Guainazzi et al. 2001).

To flesh out this modeling framework, several key questions need to
be addressed:

\noindent
1) Is the gas uplifted from the disk
surface (radiative driving by intrinsic or reprocessed disk
radiation, or a hydromagnetic wind)?

\noindent
2) If radiation pressure on atoms is efficient, what provides
the requisite shielding of the central continuum ---
A ``failed'' radiation pressure-driven wind? The
inner region of a hydromagnetic disk outflow?

\noindent
3) Is radiation pressure on dust important for the outflow dynamics?

\noindent
4) Is the BELR/BALR gas composed
primarily of ``clouds'', as has been traditionally assumed,
or can it be attributed to a continuous outflow (Murray
et al. 1995)? If clouds are present, are they coherent,
long-lived entities, and, if so, how are they confined?

In the following sections I discuss these issues in greater
detail.

\section{Radiative Driving}

Various observations have pointed to a potentially central role
for radiation pressure in driving the observed outflows. In
particular, the detection of a ``ghost of Ly$\alpha$'' (a
signature of the radiative acceleration of the N V ion by
Ly$\alpha$ flux imprinted on the C IV BAL profile) has provided
direct evidence for radiative acceleration in BALQSOs (e.g., Arav
1996). In the case of the warm-absorber component in Sy 1
galaxies, the inferred outflow speeds were shown to be consistent
with radiative driving involving the O VII
and O VIII absorption edges (e.g., Reynolds \& Fabian
1995).\footnote{Some of the edge
identifications have, however, been questioned in view of new
Chandra and XMM/Newton observations (e.g., Branduardi-Raymont et al. 2001).}
In fact, the size of the warm-abosrber
region has been deduced by comparing observed flux variability
time scales to the recombination times of the relevant oxygen
ions (e.g., Netzer et
al. 2002).\footnote{Note in this connection that, if a given ion
can be identified as dominating the acceleration $g_{\rm rad}$,
then an independent measure of the characteristic gas-injection
radius $r_0$ can be obtained from the inferred terminal speed $V_\infty$
in view of the fact that radiative acceleration typically occurs over a
scale $\Delta r \ll r_0$ (e.g.,
Arav et al. 1994). Thus $V_{\infty} \approx [2 g_{\rm rad}(r_0) r_0]^{1/2},\, g_{\rm
rad} \propto L_{\rm ion}/r^2,\,  \Rightarrow V_\infty
\propto r_0^{-1/2}.$}

Explicit models of radiatively driven disk winds in AGNs have been constructed
semianalytically (Murray et al. 1995) and
numerically (Proga et al. 2000). These models account for the
requisite shielding of the radiatively driven gas in terms of a
``failed'' inner outflow. The numerical simulations suggest that
time-dependent dynamical effects could give rise to strong
localized density enhancements and to apparently distinct
outflow components. These results are, however, subject to the
following potential caveats:

\noindent
${\bullet}$ The ability of the disk to
radiatively uplift the postulated outflow as a result of either
external or internal heating has not yet been self-consistently calculated.

\noindent
${\bullet}$ In view of the sensitive
dependence of radiative acceleration on the
irradiating spectrum, radiative transfer effects beyond simple
attenuation need to be incorporated into the numerical models
(Everett 2002 and these Proceedings).

\noindent
${\bullet}$ The radiatively driven disk
outflow model has recently been tested by HST
observations of two high-state cataclysmic variables (Hartley et
al. 2002): the strong positive correlation between UV brightness
and wind activity predicted by these models was found to be
disobeyed by both binaries, indicating that nonradiative factors
may control the mass-loss rate in these systems.

\section{Magnetic Driving}

Magnetically mediated disk outflows
(specifically, centrifugally driven winds; e.g., Blandford \&
Payne 1982) are commonly invoked in low-luminosity astrophysical
systems such as low-mass protostars (e.g., K\"onigl \& Pudritz 2000). Magnetic
driving is also believed to underlie AGN jets (e.g., Blandford
2000). An open field configuration that is compatible
with this picture has been inferred from polarization measurements in
the Galactic center circumnuclear disk (Hildebrand et al. 1993).

Magnetic driving can produce high-speed, high-momentum-discharge
disk outflows; such outflows transport angular
momentum efficiently and hence may arise naturally in accretion
disks. AGN outflows of this type would have a strongly
stratified vertical density profile and would be photoionized in their
inner regions and dusty in their outer parts. As was
demonstrated by K\"onigl \& Kartje (1994), these winds could
naturally account for the visual obscuration and UV/X-ray
attenuation inferred in Type 2 AGNs. Such winds, in fact, give
rise to effective obscuring tori that are ``fuzzy,'' as has been
inferred in objects like NGC 4151 (
Crenshaw et al. 2000). The comparatively high opacity of the dusty
regions results in the ``flattening'' (by radiation pressure) of
the ``torus'' in high-luminosity sources (a trend that has
been inferred observationally), and the reprocessing of the
central continuum radiation by the wind and disk in the
dusty outer regions by and large reproduces the measured near/mid-infrared spectra
of Seyfert galaxies. The predicted electron and dust distributions can
also account for the distinct optical/UV continuum polarization properties 
of Sy 1 and Sy 2 galaxies (Kartje 1995). Furthermore,
the magnetic pressure of these outflows provides an effective
confining agent for embedded clouds, and internal
MHD turbulence can contribute to the line
broadening exhibited by the BELR gas (Bottorff \& Ferland 2000).

As a first step toward investigating the combined effect of
magnetic and radiative driving, Everett (2002 and these Proceedings) constructed a
semianalytic model in which magnetic stresses uplift the gas from
the disk surface and the central continuum contributes radial
radiative driving further up (after the gas rises to a height
where the inner portions of the stratified outflow provide an optimal
shielding of the nuclear continuum). His model also incorporates clouds (uplifted by the
ram pressure of the continuous wind and confined by the wind
magnetic pressure) and dust. In a complementary approach, Proga (2002) carried out
exploratory numerical simulations of magnetically/radiatively
driven disk winds under a variety of (sometimes strong) simplifications.

\section{Multiphase Outflows}

As was noted in \S 1, high-resolution UV and X-ray observations
of the warm absorber gas in Sy 1 galaxies have
identified several velocity and ionization-parameter components
in each source. In the case of NGC 3783, for example,
photoionization modeling has implied a correspondence of two
distinct ionization components to a single kinematic component
(Kraemer et al. 2001), which provides strong support for the
presence of a multiphase outflow. This conclusion is supported
by the inferred smooth variation of the continuum covering
factor in the wings of the individual kinematic absorption
components in this source, which was interpreteed as pointing to a
(cloud?) substructure (Gabel et al. 2002). Although the existing
data are not yet sufficient for drawing general conclusions,
there have been indications for the coexistence of at least two
distinct ionization components also in other sources (e.g., Mrk
509; Kraemer et al. 2002b). The presence of a two-component gas
has also been suggested for the NELR of certain Seyfert galaxies
(e.g., Ogle et al. 2000; Komossa 2001).

There is also growing evidence for the existence of multiphase outflows in BALQSOs.
For example, in the case of FBQS 1044, absorption
features that, when interpreted in the context of a single-phase
model imply a distance from the central continuum source
of $\sim 0.7\, {\rm kpc}$, yield a much smaller, physically more reasonable scale of $\sim 4\,
{\rm pc}$ when reinterpreted in terms of a shielded, continuous,
low-density wind with embedded dense clouds (Everett et al. 2002
and these Proceedings). Similar apparent difficulties in other
sources (e.g., the radio-loud galaxy 3C 191; Hamann et al. 2001)
might be resolved in the same way. The inference that an obscuring
cloud is responsible for the X-ray absorption in the lensed
BALQSO UM 425 (Aldcroft \& Green, these Proceedings) is
consistent with this picture.

The notion of discrete clouds in the BELR has been challenged by
cross-correlation spectral analyses (e.g., Arav et
al. 1998; Dietrich et al. 1999), which indicate that an
implausibly high number of clouds may be required to account for
the smoothness of the measured line profiles. This conclusion
could, however, be mitigated if MHD turbulence,
and not just thermal motions, contributes to the intrinsic line
widths (Bottorff \& Ferland 2000). Discrete clouds could be entrained
into the continuous disk outflow at the disk surface (in analogy with
coronal mass ejections in the Sun). Such clouds could be
subsequently confined by the wind magnetic field (e.g.,
Everett et al. 2002; Everett 2002). An alternative possibility
is that the clouds represent transient condensations (which need
not be confined) in a turbulent medium (e.g., Bottorff \&
Ferland 2001) --- these two possibilities would have similar
spectral signatures and thus may be
hard to distinguish observationally.\footnote{Note, however, that a strong magnetic
contribution to the line widths (which implies $P_{\rm magnetic} \gg P_{\rm thermal}$ in
the emitting gas) likely corresponds to transient
density enhancements rather than to long-lived clouds (which
require pressure confinement), since magnetic confinement of
clouds is optimized when $P_{\rm magnetic,\, cloud}
\ll P_{\rm thermal,\, cloud}$ and $P_{\rm
magnetic,\, confining} \approx P_{\rm thermal,\, cloud}$.}

\section{Further Progress}
\underline{\bf Theory}
\medskip

\noindent $\bullet$ Semianalytic models have reached the
level where the basic qualitative aspects of the interplay
between magnetic and radiative driving, and the dynamics of a
multiphase outflow, can be studied. Incorporating the effect of
a disk radiation field (in addition to a central continuum) is
the next step in this effort.

\noindent $\bullet$ 2D numerical simulations, which have
revealed several suggestive features, need to be combined with a
photoionization code and generalized to 3D. The most important
questions that need to be answered are: Can the full observed
range of outflow speeds be reproduced with this mechanism? Can
radiative driving alone account for the requisite uplifting,
shielding, and acceleration for realistic AGN radiation fields
and disk models? Can the nonsteady effects identified in the
simulations account for the observed spectral characteristics?

\noindent $\bullet$ 2D and 3D MHD simulations of
centrifugally driven disk outflows have already been carried
out, but their observational implications to AGN spectra have
not yet beed studied. Ultimately, a comprehensive
radiative/magnetic driving model should be simulated with a
single code.

\medskip\noindent
\underline{\bf Observations}
\medskip

\noindent $\bullet$ Use high-resolution
optical/UV/X-ray spectroscopy to relate the distinct kinematic
and ionization components identified in warm-absorber and BALQSO
outflows; in particular, can a general case be made for a
multiphase outflow?

\noindent $\bullet$ Continue to refine the tool of
spectropolarimetry in AGN research; in particular, can the
results derived in BALQSOs be obtained also in Seyfert galaxies,
and can more evidence be uncovered for high-velocity dust?

\noindent $\bullet$ Attempt to carry out detailed
searches for a correlation between UV brightness and wind
activity
to test radiatively driven outflow models (as has recently been
done for CVs).

\noindent $\bullet$ Search for spectral
signatures of high-velocity outflows in low-$L_{\rm ion}$ sources
(e.g., BL Lac objects) in an effort to discriminate between
magnetic and radiative disk-outflow models.


\begin{references}
\reference Antonucci, R.: 1993, ARA\&A, 31, 473
\reference Arav, N.: 1996, ApJ, 465, 17
\reference Arav, N., Li, Z.-Y., \& Begelman, M.C.: 1994, ApJ,
432, 62
\reference Arav, N., et al.: 1998, MNRAS, 297, 990
\reference Blandford, R.D.: 2000, Phil. Trans. R. Soc. Lond. A,
358, 1
\reference Blandford, R.D., \& Payne, D.G.: 1982, MNRAS, 199, 883
\reference Bottorff, M.C., et al.: 1997, ApJ, 479, 200
\reference Bottorff, M.C., \& Ferland, G.: 2000, MNRAS, 316, 103
\reference Bottorff, M.C., \& Ferland, G.: 2001, MNRAS, 549, 118
\reference Brandt, W.N., Laor, A., \& Wills, B.J.: 2000, ApJ,
528, 637
\reference Branduardi-Raymont, G., et al.: 2001, A\&A, 365, 162
\reference Chiang, J., \& Murray, N.: 1996, ApJ, 466, 704
\reference Cohen, M.H., et al.: 1995, ApJ, 448, L77
\reference Crenshaw, D.M., et al.: 1999, ApJ, 516, 750
\reference Crenshaw, D.M., et al.: 2000, ApJ, 120, 1731
\reference Crenshaw, D.M., \& Kraemer, S.B.: 1999, ApJ, 521, 572
\reference Crenshaw, D.M., \& Kraemer, S.B.: 2000, ApJ, 532, L101
\reference Crenshaw, D.M., \& Kraemer, S.B.: 2001, ApJ, 562, L29
\reference Dietrich, M., et al.: 1999, A\&A, 351, 31
\reference Everett, J.: 2002, ApJ, submitted
\reference Everett, J., K\"onigl, A., \& Arav, N.: 2002, ApJ,
569, 671
\reference Gabel, J.R., et al.: 2002, ApJ, submitted
(astro-ph/0209484)
\reference Goodrich, R.W.: 1989, ApJ, 340, 190
\reference Goodrich, R.W.: 1997, ApJ, 474, 606
\reference Goodrich, R.W., \& Miller, J.S.: 1995, ApJ, 448, L73
\reference Green, P.J., et al.: 2001, ApJ, 558, 109
\reference Guainazzi, M., et al.: 2001, MNRAS, 327, 323
\reference Hamann, F.W., et al.: 2001, ApJ, 550, 142
\reference Hartley, L.E., et al.: 2002, MNRAS, 332, 127
\reference Hildebrand, R.H., et al.: 1993, ApJ, 417, 565
\reference Hough, D.H. et al.: 2002, ApJ, 123, 1258
\reference Hutchings, J.B., et al.: 2001, ApJ, 559, 173
\reference Kaastra, J.S., et al.: 2002, A\&A, 386, 427
\reference Kartje, J. F.: 1995, ApJ, 452, 565
\reference Kaspi, S., et al.: 2000, ApJ, 553, 631
\reference Kaspi, S., et al.: 2002, ApJ, 574, 643
\reference K\"onigl, A., \& Kartje, J.: 1994, ApJ, 434, 446
\reference K\"onigl, A., \& Pudritz, R.E.: 2000, in Protostars
\& Planets IV, 759
\reference Kraemer, S.B., et al.: 2002a, ApJ, 577, 98
\reference Kraemer, S.B., et al.: 2002b, ApJ, submitted (astro-ph/0208478)
\reference Murray, N., \& Chiang, J.: 1997, ApJ, 474, 91
\reference Murray, N., et al.: 1995, ApJ, 451, 498
\reference Netzer, H., et al.: 2002, ApJ, 571, 256
\reference Netzer, H., \& Laor, A.: 1993, ApJ, 404, L51
\reference Ogle, P.M., et al.: 2000, ApJ, 545, L81
\reference Proga, D.: 2002, ApJ, in press (astro-ph/0210642)
\reference Proga, D., Stone, J.M., \& Kallman, T.R.: 2000, ApJ,
543, 686
\reference Reynolds, C.S., \& Fabian, A.C.: 1995, MNRAS, 273, 1167
\reference Richards, G. T., et al.: 1999, ApJ, 513, 576
\reference Risaliti, G., Elvis, M., \& Nicastro, F.: 2002, ApJ, 571, 234 
\reference Sako, M., et al.: 2001, A\&A, 365, L168
\reference Vestergaard, M., Wilkes, B.J., \& Barthel, P.D.:
2000, ApJ, 538, L103
\end{references}
\end{document}